\documentclass[
preprint,
superscriptaddress,
nobibnotes,
amsmath,amssymb,
aps,
prm,
]{revtex4-2}

\usepackage{amssymb}
\usepackage{amsfonts}
\usepackage{amsmath}
\usepackage{graphicx}
\usepackage{dcolumn}
\usepackage{bm}
\usepackage{xcolor}
\usepackage{float}
\usepackage{textgreek}
\usepackage{upgreek}

\usepackage[colorlinks]{hyperref}
\hypersetup{
                pdfstartview={FitH},
                linkcolor=blue,
                citecolor=blue,
                filecolor=blue,
                urlcolor=blue
}

\begin{document}


\title{Atomic-layer controlled THz Spintronic emission from Epitaxially grown Two dimensional PtSe$_2$/ferromagnet heterostructures}

\author{K. Abdukayumov}
\thanks{These two authors equally contributed to the present work.}
\affiliation{Univ. Grenoble Alpes, CEA, CNRS, Grenoble INP, IRIG-Spintec, 38000 Grenoble, France}

\author{M. Mi\v cica}
\thanks{These two authors equally contributed to the present work.}
\affiliation{Laboratoire de Physique de l’Ecole Normale Sup\'erieure, ENS, Universit\'e PSL, CNRS, Sorbonne Universit\'e, Universit\'e de Paris, Paris, France}

\author{F. Ibrahim}
\affiliation{Univ. Grenoble Alpes, CEA, CNRS, Grenoble INP, IRIG-Spintec, 38000 Grenoble, France}

\author{C. Vergnaud}
\affiliation{Univ. Grenoble Alpes, CEA, CNRS, Grenoble INP, IRIG-Spintec, 38000 Grenoble, France}

\author{A. Marty}
\affiliation{Univ. Grenoble Alpes, CEA, CNRS, Grenoble INP, IRIG-Spintec, 38000 Grenoble, France}

\author{J.-Y. Veuillen}
\affiliation{Universit\'e Grenoble Alpes, CNRS, Grenoble INP, Institut NEEL, 38000 Grenoble, France}

\author{P. Mallet}
\affiliation{Universit\'e Grenoble Alpes, CNRS, Grenoble INP, Institut NEEL, 38000 Grenoble, France}

\author{I. de Moraes}
\affiliation{Univ. Grenoble Alpes, CEA, CNRS, Grenoble INP, IRIG-Spintec, 38000 Grenoble, France}

\author{D. Dosenovic}
\affiliation{Universit\'e Grenoble Alpes, CEA, IRIG-MEM, 38000 Grenoble, France}

\author{Adrien Wright}
\affiliation{Laboratoire de Physique de l’Ecole Normale Sup\'erieure, ENS, Universit\'e PSL, CNRS, Sorbonne Universit\'e, Universit\'e de Paris, Paris, France}

\author{J\'er\^ome Tignon}
\affiliation{Laboratoire de Physique de l’Ecole Normale Sup\'erieure, ENS, Universit\'e PSL, CNRS, Sorbonne Universit\'e, Universit\'e de Paris, Paris, France}

\author{Juliette Mangeney}
\affiliation{Laboratoire de Physique de l’Ecole Normale Sup\'erieure, ENS, Universit\'e PSL, CNRS, Sorbonne Universit\'e, Universit\'e de Paris, Paris, France}

\author{A. Ouerghi}
\affiliation{Universit\'e Paris-Saclay, CNRS, Centre de Nanosciences et de Nanotechnologies, 91120, Palaiseau, France}

\author{V. Renard}
\affiliation{Universit\'e Grenoble Alpes, CEA, CNRS, IRIG-PHELIQS, 38000 Grenoble, France}

\author{F. Mesple}
\affiliation{Universit\'e Grenoble Alpes, CEA, CNRS, IRIG-PHELIQS, 38000 Grenoble, France}

\author{F. Bonell}
\affiliation{Univ. Grenoble Alpes, CEA, CNRS, Grenoble INP, IRIG-Spintec, 38000 Grenoble, France}

\author{H. Okuno}
\affiliation{Universit\'e Grenoble Alpes, CEA, IRIG-MEM, 38000 Grenoble, France}

\author{M. Chshiev}
\affiliation{Univ. Grenoble Alpes, CEA, CNRS, Grenoble INP, IRIG-Spintec, 38000 Grenoble, France}

\author{J.-M. George}
\affiliation{Unit\'e Mixte de Physique, CNRS, Thales,
Universit\'e Paris-Saclay, F-91767 Palaiseau, France}

\author{H. Jaffr\`es}
\affiliation{Unit\'e Mixte de Physique, CNRS, Thales,
Universit\'e Paris-Saclay, F-91767 Palaiseau, France}

\author{S. Dhillon}
\affiliation{Laboratoire de Physique de l’Ecole Normale Sup\'erieure, ENS, Universit\'e PSL, CNRS, Sorbonne Universit\'e, Universit\'e de Paris, Paris, France}

\author{M. Jamet}
\affiliation{Univ. Grenoble Alpes, CEA, CNRS, Grenoble INP, IRIG-Spintec, 38000 Grenoble, France}

\date{\today}

\begin{abstract}

Terahertz (THz) Spintronic emitters based on ferromagnetic/metal junctions have become an important technology for the THz range, offering powerful and ultra-large spectral bandwidths. These developments have driven recent investigations of two-dimensional (2D) materials for new THz spintronic concepts. 2D materials, such as transition metal dichalcogenides (TMDs), are ideal platforms for SCC as they possess strong spin-orbit coupling (SOC) and reduced crystal symmetries. Moreover, SCC and the resulting THz emission can be tuned with the number of layers, electric field or strain. Here, epitaxially grown 1T-PtSe$_2$ and sputtered Ferromagnet (FM) heterostructures are presented as a novel THz emitter where the 1T crystal symmetry and strong SOC favor SCC. High quality of as-grown PtSe$_2$ layers is demonstrated and further FM deposition leaves the PtSe$_2$ unaffected, as evidenced with extensive characterization. Through this atomic growth control, the unique thickness dependent electronic structure of PtSe$_2$ allows the control of the THz emission by SCC. Indeed, we demonstrate the transition from the inverse Rashba-Edelstein effect in one monolayer to the inverse spin Hall effect in multilayers. This band structure flexibility makes PtSe$_2$ an ideal candidate as a THz spintronic 2D material and to explore the underlying mechanisms and engineering of the SCC for THz emission.

\end{abstract}

\maketitle


\section{Introduction}

During the past two decades, two-dimensional (2D) materials have raised tremendous interest in the scientific community for the novel and specific properties as a result of their reduced dimensionality and van der Waals character \cite{Novoselov_2004,Novoselov_2005,Cao_2018}. In particular, when isolated to one monolayer (ML), transition metal dichalcogenides (TMD) 1H-MX$_2$ (with M=Mo, W and X=S, Se) become direct bandgap semiconductors with remarkable optical properties \cite{Mak_2010,Cadiz_2017}. In addition, inversion symmetry breaking giving rise to in-plane electric field combined with strong spin-orbit coupling gives rise to two inequivalent K valleys exhibiting spin-valley locking \cite{Xiao_2012}. This property is at the origin of unique phenomena such as optical valley selection by circularly polarized light \cite{Sallen_2012}, large valley Hall and Nernst effects \cite{Mak_2014,Dau_2019}, anisotropic spin relaxation \cite{Benitez_2018} and sizeable spin-orbit torques \cite{Guimaraes_2018}. Platinum diselenide (PtSe$_2$) is a recent TMD that shows several key properties like air stability \cite{Zhao_2017}, high carrier mobility \cite{Bonell_2022}, high photoelectrical response in the near-infrared range \cite{Cao_2021} and defect-induced ferromagnetism \cite{Avsar_2020}. By selectively substituting selenium by sulfur atoms, the polar SePtS Janus material could also be synthesized \cite{Sant_2020}. Moreover, it exhibits a layer-dependent bandgap: one monolayer of PtSe$_2$ is a ~1.9 eV bandgap semiconductor while for layers greater than 3 ML it becomes semi-metallic \cite{Villaos_2019}. Finally, owing to the large atomic weight of platinum, PtSe$_2$ possesses the greatest spin-orbit coupling among transition metal diselenides making it an excellent candidate to study spin-to-charge conversion (SCC) in van der Waals materials. In this respect, it constitutes a unique system to observe the transition from inverse Rashba-Edelstein effect (IREE) in the semiconducting phase to inverse spin Hall effect (ISHE) in the semi-metallic phase. Furthermore, 1T-PtSe$_2$ shows crystal inversion symmetry favoring in-plane spin textures at interfaces whereas spins are mostly locked out-of-plane in 1H-MX$_2$ preventing SCC for in-plane polarized spin currents. Angle and spin-resolved photoemission spectroscopy measurements have indeed demonstrated the existence of in-plane helical spin textures in the valence band of monolayer PtSe$_2$ \cite{Yao_2016,Yan_2017}. To study SCC in 2D materials, spin pumping is typically used and has, for example, been used to measure the strength of IREE \cite{Bangar_2022}. On the other hand, Terahertz (THz) emission spectroscopy has become a powerful and established tool to probe SCC \cite{Seifert_2016,Rongione_2023}. The associated devices are known as spintronic THz emitters and were first demonstrated with metallic ferromagnetic (Co, Fe…)/nonmagnetic (Pt, W…) thin films \cite{Hawecker_2021,Dang_2020,Cheng_2021}. They present several advantages compared to other THz sources such as broadband THz emission, high efficiency and easy control of radiation parameters. To date, very few 2D materials have been incorporated in THz spintronic emitters \cite{Cheng_2019,Nadvornik_2022,Khusyainov_2021,Cong_2021} and they all exhibit the 1H crystal structure which is not favorable to convert in-plane polarized spins into charge currents. Indeed, the combination of inversion symmetry breaking and strong SOC forces the spin of electrons with finite momentum to be out-of-plane.
In this work, we have grown single crystalline mono and multilayers of 1T-PtSe$_2$ on graphene by van der Waals epitaxy \cite{Dau_2018,Dau_2019b,Vergnaud_2020,Velez_2022} and developed a soft sputtering process to grow amorphous CoFeB on top of PtSe$_2$ obtaining atomically sharp CoFeB/PtSe$_2$ interface. We used a full set of characterization tools to demonstrate the structural and chemical preservation of PtSe$_2$ after CoFeB deposition. SCC was then studied on these advanced 2D samples as a function of PtSe$_2$ thickness from 1 to 15 ML using THz emission spectroscopy that showed the efficient generation of THz electric fields through SCC. This THz emission is shown to arise from the 1T crystal structure and large spin-orbit coupling of PtSe$_2$. Indeed, negligible signals were obtained for 1T-VSe$_2$ with low spin-orbit coupling and for 1H-WSe$_2$ with 1H crystal structure. Further, the THz electric field clearly shows a two-step dependence on the PtSe$_2$ thickness that we interpret as the transition from the IREE in the semiconducting regime to the ISHE in the semimetallic regime. As shown by ab initio calculations, the IREE arises from the large Rashba spin splitting at the PtSe$_2$/graphene interface by the combination of large spin-orbit coupling and the interfacial electric dipole. By fitting the thickness dependence, we could extract the out-of-plane spin diffusion length in PtSe$_2$ to be 2-3 nm and found that SCC by IREE at the PtSe$_2$/Gr interface is twice as efficient than that of ISHE in bulk PtSe$_2$. For its unique thickness dependent electronic structure, PtSe$_2$ spintronic structures enable to observe the transition from IREE to ISHE and opens up perspectives for the application of these structures to THz spintronics. Furthermore, by adjusting the Fermi level position via gating the material, it will be possible to modulate and dramatically enhance the SCC efficiency and achieve further large enhancement of the spintronic THz emission. 

\section{Results and discussion}

\subsection{Sample growth and characterization}

\begin{figure}
  \includegraphics[width=\linewidth]{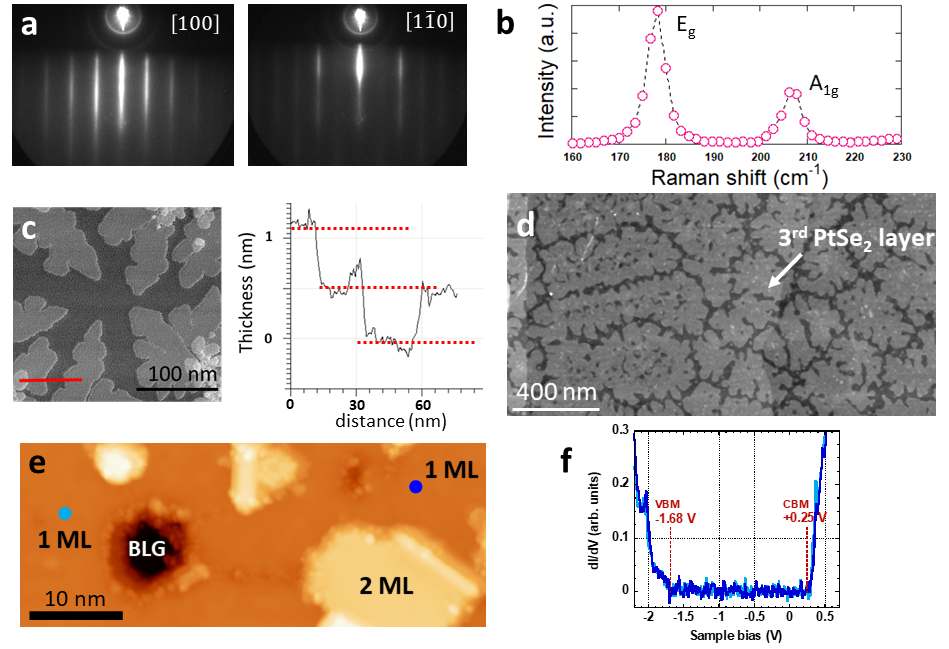}
  \caption{(a) RHEED patterns of 10 ML PtSe$_2$ grown on Gr/SiC recorded along two different azimuths [100] and [1$\bar{1}$0]. (b) Raman spectrum on the same sample. (c), (d) AFM images for $<$1 ML and 3 layers of PtSe$_2$ respectively. In (c), the height profile along the red solid line is shown on the right. (e) STM image showing a portion of uncovered bilayer graphene substrate (labelled BLG), a few domains of the monolayer 1T-PtSe$_2$ phase (labelled 1 ML) and an island of bilayer (labelled 2 ML). Image size: 50$\times$20 nm$^2$, sample bias: +1.20 V. (f) STS spectra taken on two different 1 ML domains, at locations indicated by the blue dots of the same color in (e). The setpoint voltage (current) is +1.2 V (0.5 nA). The curves represent the lock-in signal. VBM, E$_F$ and CBM correspond to the valence band maximum, the Fermi level and the conduction band minimum, respectively.
}
  \label{Fig1}
\end{figure}

All the samples are grown under ultra-high vacuum (UHV) conditions in the low 10$^{-10}$ mbar pressure range. Substrates are epitaxial graphene on undoped SiC(0001) thermally prepared by surface graphitization \cite{Kumar_2016,Pallecchi_2014}. The low spin-orbit interaction in graphene and SiC ensures negligible SCC. Prior to the growth, the substrates are annealed at 800°C during 30 minutes under UHV to desorb contaminants. The substrate is then maintained at 300°C during the growth. Platinum is evaporated using an electron gun at a constant rate of 0.003125 \AA/s for 1 ML PtSe$_2$ and 0.00625 \AA/s for the other thicknesses as monitored by a quartz microbalance. Selenium is evaporated using a Knudsen cell with a vapor pressure of 10$^{-6}$ mbar at the sample position. This ensures a large Se:Pt ratio greater than 10 necessary to avoid the formation of Se vacancies in the material. Up to 3 ML of PtSe$_2$, the growth is performed in this one step whilst for thicker films, it is achieved in several steps alternating deposition and annealing at 700°C under Se to smooth out the surface. At the end of the growth, all the films are annealed at 700°C during 15 minutes to improve the crystal quality. PtSe$_2$ grows epitaxially on graphene and typical reflection high energy electron diffraction patterns (RHEED) are shown in Fig. 1a along two azimuths 30° apart from each other. The observed thin streaks and anisotropic character demonstrate the high crystalline quality of PtSe$_2$ films, \textit{i.e.} large grain size and good crystalline orientation. As usually observed in van der Waals epitaxy of 2D materials on graphene, the mosaic spread measured by azimuthal in-plane x-ray diffraction shows a maximum of $\pm$5° \cite{Dau_2018}. Thanks to the air stability of PtSe$_2$, it is possible to perform ex-situ characterization. The Raman spectrum (see Experimental section) for 10 ML of PtSe$_2$ is shown in Fig.~\ref{Fig1}b. We clearly distinguish the two characteristic vibration modes of PtSe$_2$ E$_g$ (in-plane) at 177.9 cm$^{-1}$ and A$_{1g}$ (out-of-plane) at 206.9 cm$^{-1}$ with typical full width at half maximum (FWHM) of 4 cm$^{-1}$ \cite{ElSachat_2022}. Atomic force microscopy (AFM) images performed on less than 1 and 3 ML are shown in Fig.~\ref{Fig1}c and ~\ref{Fig1}d, respectively. For partially covered graphene in Fig.~\ref{Fig1}c, the height profile gives identical PtSe$_2$/Gr and PtSe$_2$/PtSe$_2$ step heights of $\approx$0.5 nm corresponding to the c-lattice parameter of bulk PtSe$_2$. Moreover, for 3 ML in Fig.~\ref{Fig1}d, the AFM image illustrates the quasi completion (96.6 \%) of the third layer and the layer-by-layer film growth. This growth mode allows for the fine control of the number of deposited PtSe$_2$ monolayers. 

Scanning tunnelling microscopy (STM) and spectroscopy (STS) measurements were made on a sample grown on bilayer graphene (BLG). The nominal thickness of the PtSe$_2$ film was 1 ML. As shown in Fig.~\ref{Fig1}e, the sample consists mostly of the monolayer (1 ML) PtSe$_2$ , with small islands of bilayer PtSe$_2$ (2 ML) and patches of the bare BLG substrate. The lateral size of the 1 ML islands is in the ten nanometers range. Fig.~\ref{Fig1}e is a medium scale STM image of an area with different semiconducting objects: a few connected 1 ML PtSe$_2$ grains and a 2 ML island. Larger scale STM image and height distribution are shown in the Supp. Information. Fig.~\ref{Fig1}f displays spectra taken on two 1 ML grains. The curves are quite similar (this also holds for the spectra taken on the other grains), and reveal a gap estimated to be 1.93 eV. The Fermi level E$_F$ (corresponding to zero sample bias) is close to the conduction band minimum (CBM), located at +0.25 V, whereas the valence band maximum (VBM) is found to be -1.68 eV below E$_F$. From a series of measurements on 1 ML islands on a BLG substrate we find an average gap width of 1.95$\pm$0.05 eV, with a CBM (VBM) located 0.24$\pm$0.03 eV above (1.70$\pm$0.05 eV below) the Fermi level. This gap value is similar to the ones reported in previous papers for 1 ML PtSe$_2$ grown on Highly Ordered Pyrolytic Graphite (HOPG), which ranged from 1.8 eV \cite{Jingfeng_2021} to 2.00$\pm$0.10 eV \cite{Zhang_2021} and 2.09 eV \cite{Wu_2021}. In these reports, the CBM of the 1 ML PtSe$_2$ on HOPG substrate is located typically 0.55 eV above the Fermi level, a value larger than the one we find here for a BLG substrate. This shift of the CBM is consistent with the difference in the work functions of BLG and HOPG, which amounts to approximately 0.20 eV \cite{Hibino_2009}, as expected for a weak Fermi level pinning at the graphene/TMD interface \cite{Liu_2016}. Following the same argument, the CBM of 1 ML PtSe$_2$ grown on monolayer graphene on SiC(0001) as the one used for THz emission should be located between 100 and 150 meV above the Fermi level. It results from the decrease by 100 meV \cite{Hibino_2009} to 135 meV \cite{Filleter_2008} of the substrate work function between monolayer and bilayer graphene on SiC(0001). This statement is supported by experimental data for other semiconducting TMD grown on epitaxial graphene \cite{LeQuang_2017}. 

\begin{figure}
  \includegraphics[width=\linewidth]{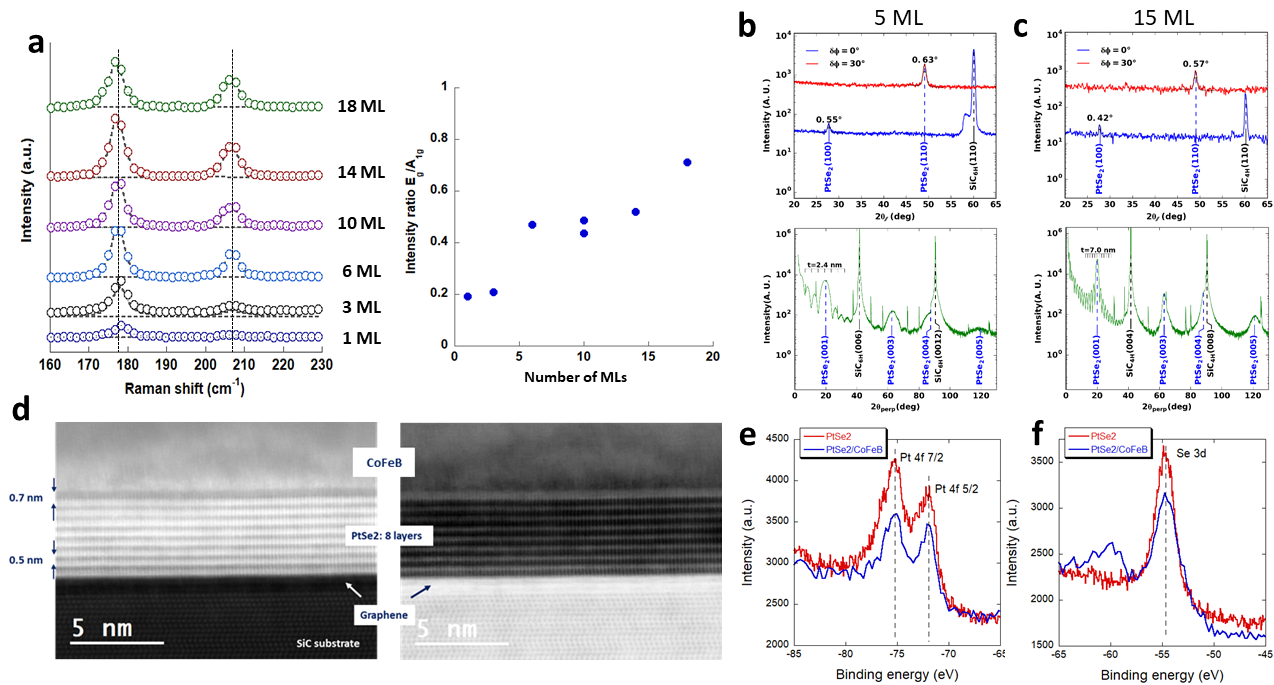}
  \caption{(a) Raman spectra of Al/CoFeB/PtSe$_2$/Gr/SiC samples for different thicknesses of PtSe$_2$. The vertical dashed black lines correspond to the peak positions of Fig.~\ref{Fig1}b. On the right, the intensity ratio E$_g$/A$_{1g}$ plotted as a function of the PtSe$_2$ thickness. (b), (c) Top: in-plane diffraction radial scans for 5 ML and 15 ML of PtSe$_2$ respectively. The scans were performed along 2 directions  at 0° and 30° of the SiC(110) reciprocal direction. The full width at half maximum is shown above each PtSe$_2$ peak. The grazing incidence was set to an angle 0.5° and 0.6° respectively to maximize the intensity of the PtSe$_2$(110) peak. The thickness of the PtSe$_2$ layer and presence of CoFeB overlayer result in an optimum incidence angle slightly above the critical angle of the substrate.  Bottom: specular $\theta$/2$\theta$ x-ray diffraction scans showing PtSe$_2$(00L) peaks, 2 major SiC peaks and many other weak intensity peaks related to the SiC polytype 4H or 6H. In addition, weak intensity peaks due to K$_\beta$ radiation are not completely removed by the multilayer mirror and the K$_\beta$ filter. At low angles, clearly visible fringes around the PtSe$_2$(001) Bragg peak show the quality of the PtSe$_2$ layer and allow to estimate its thickness reported above. (d) STEM/HAADF images in cross section of 8 ML of PtSe$_2$ covered with CoFeB in bright field (left) and dark field (right). Each PtSe$_2$ layer can be individually identified and a 0.7 nm gap is present at the CoFeB/PtSe$_2$ interface. (e), (f) XPS spectra of Pt 4f and Se 3d core levels respectively. Red (blue) before (after) CoFeB coverage.}
  \label{Fig2}
\end{figure}

We conclude that the Fermi level of 1 ML PtSe$_2$ is systematically shifted towards the CBM as a consequence of electron transfer from graphene to PtSe$_2$ \cite{Dappe_2020,Dau_2018}. This charge transfer is discussed later from the theoretical point of view and for its consequence on SCC. 
Finally, in-plane and out-of-plane x-ray diffraction (see Experimental section) on 6 ML PtSe$_2$ (not shown) yield the following lattice parameters: a=3.703 \AA\ and c=5.14 \AA. It corresponds to an in-plane compression of -0.6 \% and out-of-plane expansion of +1.1 \% with respect to bulk PtSe$_2$ \cite{Furuseth_1965}. The full width at half maximum (FWHM) of (100) and (110) Bragg peaks are 0.57° and 0.61° , respectively, close to the instrumental resolution. It reveals the large grain size of the PtSe$_2$ film.\\
Next, in order to perform THz spintronic emission, we deposit 3 nm of amorphous CoFeB on top of PtSe$_2$ covered by 4 nm of aluminum that transforms into AlO$_x$ to protect the ferromagnetic layer against oxidation. We use soft sputtering conditions for CoFeB deposition not to damage the PtSe$_2$ surface. We set the argon pressure to 1.25$\times$10$^{-2}$ mbar and the deposition rates are 0.319 \AA/s for CoFeB and 0.265 \AA/s for Al respectively. Moreover, the MBE evaporation chamber being connected to the magnetron sputtering reactor, all PtSe$_2$ films are transferred in-situ under UHV conditions fully preserving the PtSe$_2$ surface from contamination. In Fig.~\ref{Fig2}, we used a full set of characterization tools to study the impact of CoFeB deposition on PtSe$_2$ properties. In Fig.~\ref{Fig2}a, all the Raman spectra for 1-18 ML of PtSe$_2$ exhibit the two same E$_g$ and A$_{1g}$ vibration modes demonstrating that the TMD film retains its crystal integrity after CoFeB deposition. For 10 ML of PtSe$_2$, E$_g$ and A$_{1g}$ peaks fall exactly at the same positions with and without CoFeB. We observe an increase of the E$_g$/A$_{1g}$ intensity ratio when increasing the thickness as already observed for pristine PtSe$_2$ \cite{Yan_2017}. This result is confirmed by in-plane and out-of-plane x-ray diffraction as shown in Fig.~\ref{Fig2}b and ~\ref{Fig2}c for 5 ML and 15 ML of PtSe$_2$, respectively. First, the radial scans along the reciprocal directions at 0° and 30° of the substrate SiC(hh0) directions show the diffraction peaks of the substrate and PtSe$_2$. The anisotropic character indicates that the layers are crystallographically well oriented in epitaxy with respect to the substrate. From the Bragg peak positions, we obtain the in-plane and out-of-plane lattice parameters: a=3.709 \AA\ (resp. a=3.713 \AA) and c=5.18 \AA\ (resp. c=5.14 \AA) for 5 ML (resp. 15 ML). They perfectly match the values obtained on 6 ML of PtSe$_2$ without CoFeB. Moreover, the FWHM of in-plane (100) and (110) Bragg peaks (0.55°, 0.63° for 5 ML and 0.42°, 0.57° for 15 ML) are comparable to that of PtSe$_2$ without CoFeB and confirms the large grain size. In addition, from the diffraction fringes around (001) Bragg peaks, we can conclude that the CoFeB/PtSe$_2$ interface is flat at the atomic scale and deduce the film thickness to be 2.4 nm (2.5 nm expected) for 5 ML and 7.0 nm (expected 7.5 nm) for 15 ML. 

These results confirm that PtSe$_2$ films are not affected by the deposition of CoFeB.  The interface atomic structure and chemical composition are further investigated by scanning transmission electron microscopy (STEM) in the high angular annular dark field (HAADF) mode and in situ x-ray photoemission spectroscopy (XPS) respectively. Fig.~\ref{Fig2}d shows in the dark and bright field modes 8 ML of PtSe$_2$ epitaxially grown on Gr/SiC and covered by CoFeB. We clearly distinguish an atomically sharp interface with a 0.7 nm gap between the PtSe$_2$ and CoFeB. In Fig.~\ref{Fig2}e and ~\ref{Fig2}f, XPS spectra of Pt 4f and Se 3d core levels are recorded on 3 ML of PtSe$_2$ before and after deposition of 1 nm of CoFeB. For both elements, the spectra superimpose before and after CoFeB deposition while the signal decrease after CoFeB deposition is due to the partial absorption of photoelectrons by the metallic film. To summarize, we can conclude from TEM and XPS analysis that the PtSe$_2$ surface is preserved after CoFeB deposition.

\subsection{THz spintronic emission}

\begin{figure}
  \includegraphics[width=\linewidth]{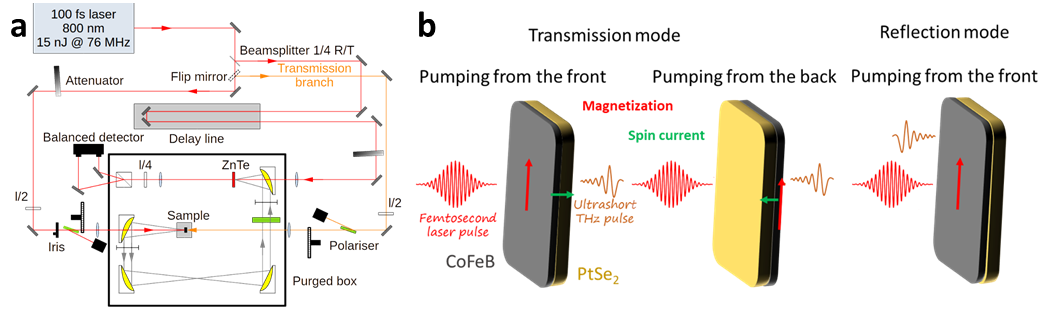}
  \caption{(a) Schematics of the THz setup. (b) The 3 different measurement geometries: transmission mode with pumping from the front and the back sides and reflection mode. CoFeB is in grey and PtSe$_2$ in yellow. The red (green) arrow corresponds to the CoFeB magnetization (injected spin current). }
  \label{Fig3}
\end{figure}

\begin{figure}
  \includegraphics[width=\linewidth]{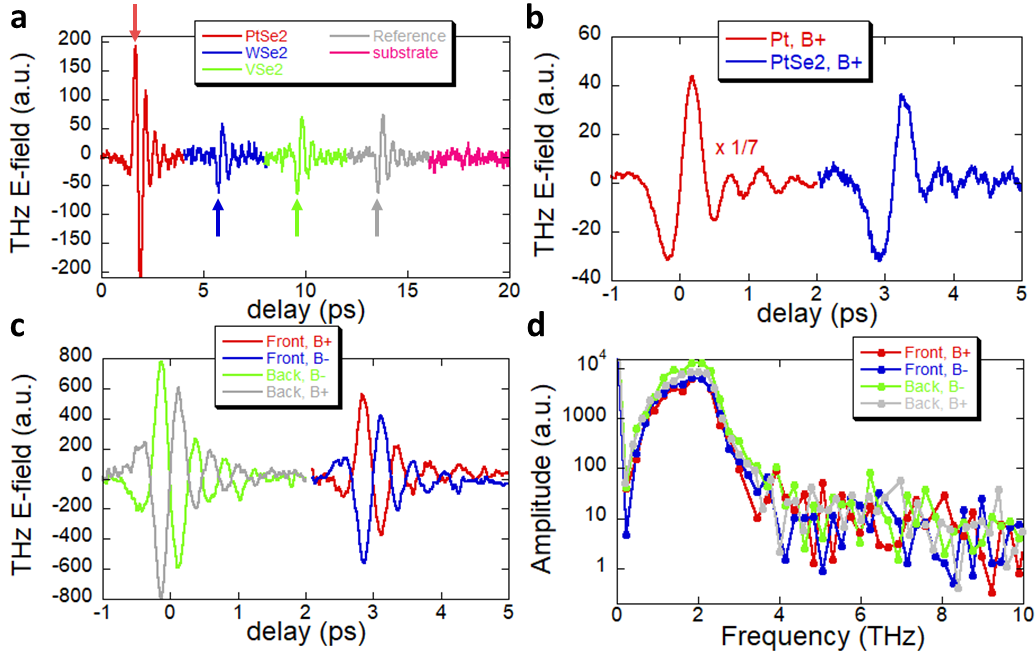}
  \caption{(a) THz E-field emitted by CoFeB/10 ML PtSe$_2$ (red), CoFeB/10 ML WSe$_2$ (blue), CoFeB/10 ML VSe$_2$ (green), CoFeB/Gr reference sample (grey) and bare Gr/SiC substrate (pink). The arrows indicate the position of the first THz E-field peak: positive for PtSe$_2$ and negative for WSe$_2$, VSe$_2$ and reference sample. All the measurements were carried out in reflection mode with a 15 fs laser pulse and 0.1 mm ZnTe detector and with a fixed 20 mT (B+) applied magnetic field. (b) THz emission for 5 nm of Pt and 5 nm (10 ML) of PtSe$_2$ measured in the same conditions and magnetic field. They show the same sign but the Pt signal is 7 times larger than the one of PtSe$_2$. The measurements were carried out in transmission mode from the backside with a 100 fs laser pulse and a 2 mm ZnTe detector. (c) THz emission of 10 ML PtSe$_2$ measured in transmission mode with a 15 fs laser pulse and 1 mm ZnTe detector. The sample was pumped from the back (front) side with positive magnetic field in grey (red) and negative magnetic field in green (blue). (d) Fourier transform of the four previous spectra.  }
  \label{Fig4}
\end{figure}

Measurements of the emitted THz waves from CoFeB/TMD samples are performed using THz emission time domain spectroscopy (TDS) as depicted in Fig.~\ref{Fig3}a. The sample is placed in a magnetic mount to apply an in-plane magnetic field of approximately 20 mT, with independent rotation of the sample and magnetic field. It is excited by a Ti:Sapphire oscillator (Coherent MIRA, 15-100 fs pulse length and horizontal polarization) at a wavelength of 800 nm in the near infrared (NIR) range. The average power exciting the sample is about 300 mW (after modulation by a chopper), adjustable by a half-wave plate followed by a polarizer. The emitted radiation is collected by a system of parabolic mirrors and the residual NIR pump is filtered by a Teflon plate. The last parabolic mirror with a hole in the center superimposes the emitted THz and the probe NIR pulse and focuses the THz beam on a ZnTe crystal for electro-optic detection with sensitivity set to horizontal component of the THz electric field (E-field). The ultrashort femtosecond laser pulse excites spin polarized electron-hole pairs in the ferromagnet (CoFeB) creating a net spin accumulation that generates a spin current into the non-magnetic layer (here the TMD) to be converted into a transverse charge current by ultrafast SCC. This results in an emitted electromagnetic pulse with frequencies in the THz range. As shown in Fig.~\ref{Fig3}b, we use both transmission (from the front and back sides) and reflection modes, which are used to show the spintronic origins of the THz emission. 
In Fig.~\ref{Fig4}a, we first compare spintronic THz emission from different samples: Gr/SiC bare substrate, CoFeB/Gr/SiC reference sample, CoFeB/VSe$_2$, CoFeB/WSe$_2$ and CoFeB/PtSe$_2$. VSe$_2$, WSe$_2$ and PtSe$_2$ are all 10 ML-thick. The epitaxial growth of VSe$_2$ and WSe$_2$ on graphene are detailed in \cite{Velez_2022} and \cite{Dau_2019c}, respectively. First, we clearly see that only CoFeB/PtSe$_2$ gives an enhanced THz signal with respect to the reference sample. Moreover, the Gr/SiC substrate adds no background signal. Next, CoFeB/VSe$_2$, CoFeB/WSe$_2$ and the reference sample exhibit the same THz emitted signal in sign and amplitude. We thus conclude that the origin of SCC is the same in all three samples. The common feature between the three samples being the AlO$_x$/CoFeB interface, we believe that the THz signal comes from the SCC at this interface through the inverse Rashba-Edelstein mechanism and/or from the self-emission of CoFeB due to unbalanced spin-flips at both interfaces. It also demonstrates that SCC in Gr/SiC, VSe$_2$ and WSe$_2$ is negligible. This can be justified by the low spin-orbit coupling of Gr/SiC and 1T-VSe$_2$ whereas spin-valley locking favors out-of-plane spin polarization in 2H-WSe$_2$ which partially prevents SCC when in-plane spins are injected. By looking at the first E-field maximum of each sample (indicated by arrows in Fig.~\ref{Fig4}a), we notice that it is opposite for PtSe$_2$ pointing out another SCC mechanism taking place in PtSe$_2$, opposite to the Rashba-Edelstein effect at the AlO$_x$/CoFeB interface. In the following, we focus on THz spintronic emission from PtSe$_2$. In Fig.~\ref{Fig4}b, we first compare the THz signals for pure Pt (5 nm) and PtSe$_2$ (10 ML, 5 nm) with the magnetic field applied in the same direction (B+). The SCC in Pt and PtSe$_2$ have the same sign and the magnitude is 7 times higher in Pt. Then, to verify the magnetic character of THz emission, we apply the magnetic field in two opposite directions (B+ and B-) and record the THz E-field in transmission mode by optically pumping the sample from the front and the back sides as shown in Fig.~\ref{Fig4}c. The opposite signs for B+ and B- and between front and back side pumping indicate the magnetic nature of the emission due to SCC in PtSe$_2$. The overall signal being less when pumping from the front side, it proves that SiC partially absorbs THz waves. Finally, in Fig.~\ref{Fig4}d, we plot the Fourier transform of the 4 spectra shown in Fig.~\ref{Fig4}c and find the same broad band emission in the 0-4 THz range which is comparable to other THz spintronic emitters \cite{Wang_2022}. In order to study the influence of the ferromagnetic layer and deposition technique on THz emission, we performed the same measurements using 3 nm of cobalt deposited in situ on 10 ML of PtSe$_2$ by electron beam evaporation (see Supp. Info.). We find very similar results showing that the ferromagnetic layer and deposition method have almost no influence on THz emission from PtSe$_2$.  

We then study the magnetic field angle ($\theta$) dependence in the geometry shown in Fig.~\ref{Fig5}a where a horizontal linear polarizer is introduced between the sample and the detector. Along a full 0°-360° rotation as shown in Fig.~\ref{Fig5}b, we obtain a typical cos($\theta$) emission pattern with a phase reversal for opposite magnetic field directions (B+ at 0° and B- at 180°) which is a clear experimental evidence of SCC process in PtSe$_2$. No sizeable asymmetry in the emission lobes is observed meaning that non-magnetic contributions to the THz signal are negligible in this material. Regarding the sample angle ($\phi$) dependence in Fig.~\ref{Fig5}c, we observe an isotropic signal with respect to the azimuthal crystalline orientation as expected for SCC like ISHE or IREE \cite{Rongione_2022}. Moreover, the linear dependence of the THz signal (E$_{B+}$+E$_{B-}$)/2 on the laser pump  power in Fig.~\ref{Fig5}d is another experimental evidence of SCC in PtSe$_2$. 

\subsection{SCC mechanism in PtSe$_2$}

\begin{figure}
  \includegraphics[width=0.8\linewidth]{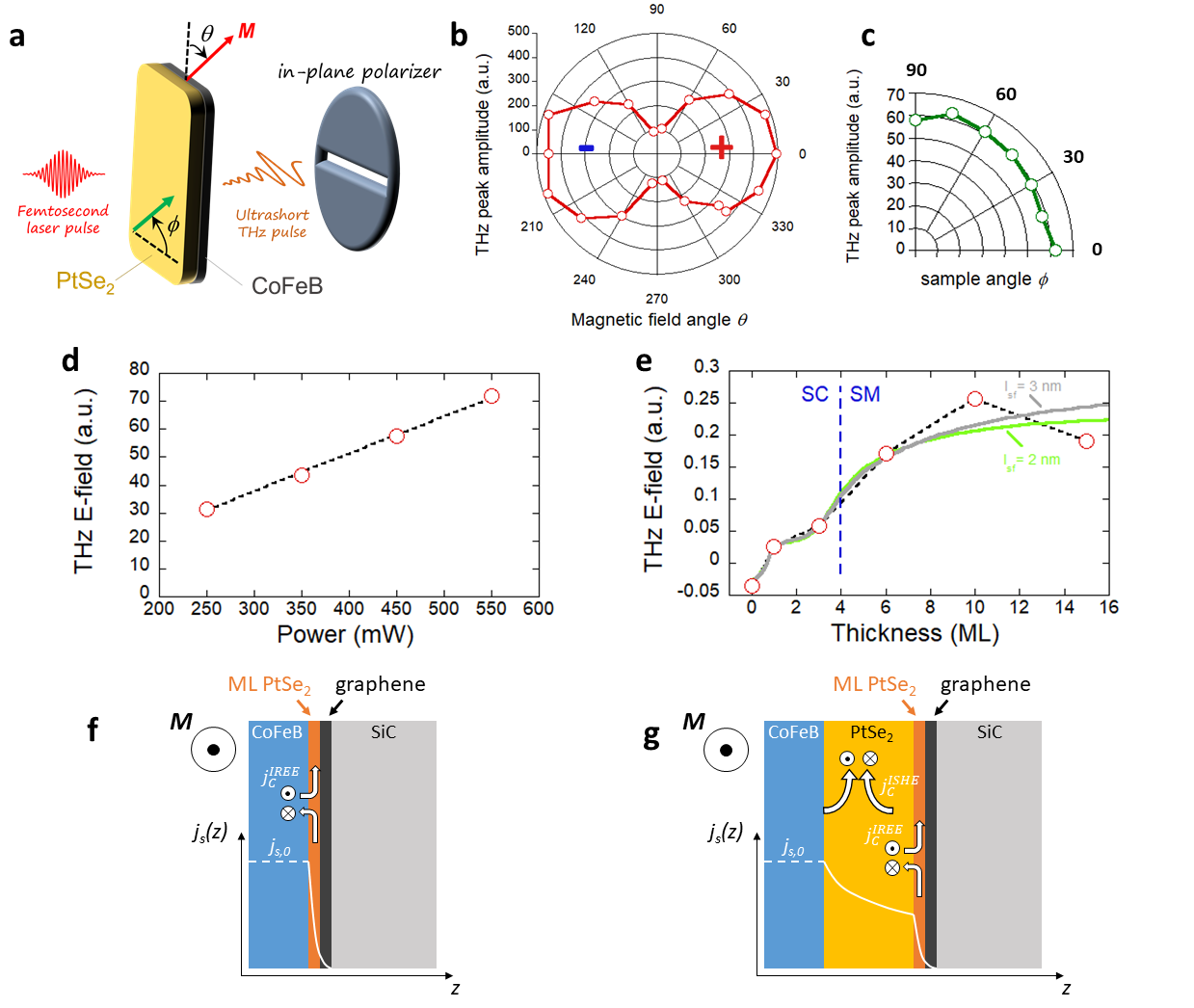}
  \caption{(a) Geometry used to measure THz emission angular dependence. The polarizer selects only in-plane polarized THz waves. $\theta$ is the angle between the vertical axis and the applied magnetic field of 20 mT. The magnetization $M$ of CoFeB is along the field direction because of its soft magnetic character. $\phi$ is the angle between the sample and horizontal axis. (b) $\theta$ dependence of the THz peak amplitude. The $+$ and $-$ signs correspond to the phase of the THz E-field. The sample was fixed and the measurements were performed in reflection mode with a laser pulse of 15 fs and a 0.1 mm ZnTe detector.
(c) $\phi$ dependence of the THz peak amplitude. The magnetic field was fixed. (d) Laser power dependence of $\frac{E_{B+}+E_{B-}}{2}$ (red open circles). The measurements were performed in transmission pumped from the backside with a laser pulse of 100 fs and a 2 mm ZnTe detector. (e) Thickness dependence of $\frac{E_{B+}+E_{B-}}{2}$ (red open circles) after normalization by laser and THz absorption by PtSe$_2$ (see Supp. Info.). The measurements were performed in transmission pumped from the backside with a laser pulse of 100 fs and a 2 mm ZnTe detector. The grey (green) solid lines are fits to the experimental data using Eq.~\ref{eq1} and $l_{sf}$=3 nm ($l_{sf}$=2 nm). (f) and (g) Schematics of the SCC in PtSe$_2$ from one layer (IREE) to multilayers (IREE+ISHE) respectively. White curve: profile of $j_s(z)$. $j_s(0)$ is the effective spin current injected in PtSe$_2$ from optically pumped CoFeB.}
  \label{Fig5}
\end{figure}

\begin{figure}
  \includegraphics[width=\linewidth]{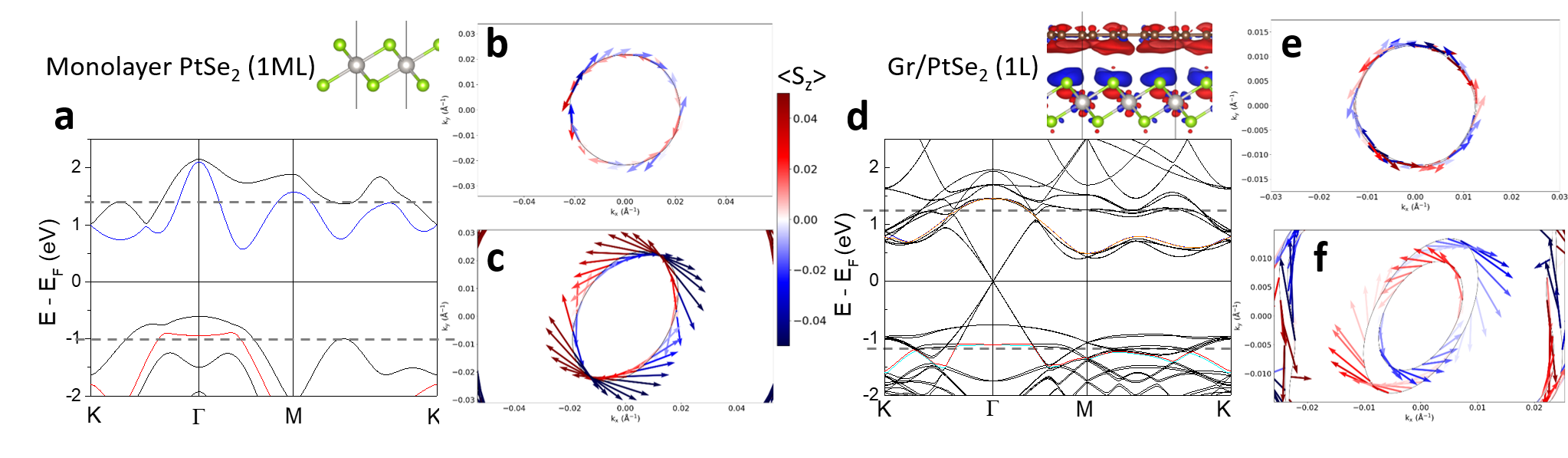}
  \caption{Rashba effect at PtSe$_2$/Gr interface. (a) Band structure of a freestanding PtSe$_2$ monolayer calculated by including spin-orbit coupling. In the shown crystal structure, the grey (green) balls represent the Pt (Se) atoms, respectively. (b) Calculated two-dimensional spin textures for the blue CB band in (a) at an energy cut $E = E_F + 0.4$ eV corresponding to the black dashed line. (c) Spin textures for the red VB in (a) at an energy cut of $E = E_F - 1$ eV. The arrows in the spin texture plots represent the $S_x$ and $S_y$ spin projections while the colour code corresponds to the $S_z$ component. (d) Calculated band structure of monolayer-PtSe$_2$/Gr heterostructure. The charge transfer occurring at the interface can be depicted on the crystal structure by red (blue) clouds corresponding to charge depletion (accumulation) taken at an isosurface value of 10$^{-4}$ $e$/\AA$^3$. (e, f) Spin textures in PtSe$_2$/Gr heterostructure calculated for the CB (at an energy cut $E = E_F + 1.25$ eV) and VB (at an energy cut $E = E_F – 1.1$ eV) chosen as in (b, c). The Rashba spin plitting is clearly visible in the VB.}
  \label{Fig6}
\end{figure}

In order to study the origin of SCC in PtSe$_2$, we performed thickness dependent measurements as shown in Fig.~\ref{Fig5}e. For this, we pumped the magnetic layer from the backside and measured the THz signal in transmission mode. We took into account the absorption of the laser and THz waves by PtSe$_2$ to normalize the signal (see Supp. Info.). The curve exhibits a two-step behavior that we interpret as IREE for thin PtSe$_2$ coverage (first step) and ISHE for thicker PtSe$_2$ (second step). Indeed, the second step appears at the semiconductor (SC) to semimetallic (SM) transition of PtSe$_2$ as given by electrical van der Pauw measurements (see Supp. Info.). This is illustrated in Fig.~\ref{Fig5}f and ~\ref{Fig5}g respectively where we assume that IREE takes place in the PtSe$_2$ monolayer in contact with graphene and ISHE in the semi metallic bulk PtSe$_2$. Based on these assumptions, we use a simple spin diffusion model to fit the experimental data. First, we assume that the spin current generated by optical excitation in CoFeB is completely absorbed in PtSe$_2$ (spin sink model). We denote $d_N$, $l_{sf}$, $j_{s0}$, $\theta_N$ and $\theta_I$ as the PtSe$_2$ thickness, spin diffusion length in PtSe$_2$, the spin current at the CoFeB/PtSe$_2$ interface, the spin Hall angle of PtSe$_2$ and effective spin Hall angle at the PtSe$_2$/Gr interface, respectively. We use the following equations:

\begin{equation}
\begin{split}
    j_s(z)/j_s(0) &=exp(-z/l_{sf}) \\
    J_N(d_N)/j_s(0) &=\int_{0}^{d_N} j_s(z)/j_s(0) dz = l_{sf}(1-exp(-d_N/l_{sf})
    \end{split}
    \label{eq1}
\end{equation}

to describe the spin current $j_s(z)$ at a depth $z$ from the CoFeB/PtSe$_2$ interface and the layer-integrated spin current in PtSe$_2$ $J_N(d_N)$. The profile of $j_s(z)$ is depicted in Fig.~\ref{Fig5}f and ~\ref{Fig5}g. The total charge current generated by IREE and ISHE can be written as:
$I_c=H(z-d_N^I)\theta_I j_s(d_N) d_N^I+H(z-d_N^C) \theta_N J_N(d_N)$ where $H$ is a step function and $d_N^I$ (resp. $d_N^C$) the PtSe$_2$ thickness for which IREE (resp. ISHE) sets in. The Rashba-Edelstein length is then defined as: $\lambda_{IREE}=\theta_I d_N^I$ and we assume $\theta_I$ to be independent of $d_N$. The best fit to the experimental data gives the following parameters: $d_N^I$=0.35 nm; $d_N^C$=1.7 nm; $l_{sf}$ between 2 nm (green fitting curve in Fig.~\ref{Fig5}e) and 3 nm (grey fitting curve in Fig.~\ref{Fig5}e) and $\theta_I/\theta_N\approx$2. $d_N^I$ is comparable to 1 ML of PtSe$_2$ (0.27 nm) in agreement with IREE and $d_N^C$ corresponds to 3-4 ML of PtSe$_2$ at the transition from semiconductor to semimetal in good agreement with ISHE. We find a short spin diffusion length between 2 nm (4 ML) and 3 nm (6 ML). The equivalent platinum thickness (corresponding to the Pt amount in 2 nm and 3 nm of PtSe$_2$) is between 0.65 nm and 0.975 nm demonstrating that the spin flip rate with flow along the normal to the film plane is larger in PtSe$_2$ than in Pt. This finding raises the question of spin diffusion with flow along the normal in lamellar systems. In particular, the role of van der Waals gaps needs to be taken into account. The short spin diffusion length could be explained if vdW gaps act as effective tunnel barriers in spin transport: during the tunnelling time between two PtSe$_2$ layers, in-plane spin flips occur within individual PtSe$_2$ layers limiting the vertical spin transport. This would deserve further investigation and the development of a more general theory which is out of the scope of the present work. Finally, we find that IREE and ISHE have the same sign and IREE is twice more efficient than ISHE. To support our conclusions and understand the mechanisms responsible for the THz spintronic emission, we have performed first principles calculations based on the projector-augmented wave (PAW) method \cite{Blochl_1994} as implemented in the VASP package \cite{Kresse_1993,Kresse_1996,Kresse_1996b} using the generalized gradient approximation \cite{Perdew_1996} and including spin-orbit coupling. The PtSe$_2$/Gr heterostructure was constructed by matching 2$\times$2 supercell of 1T-PtSe$_2$ with 3$\times$3 supercell of graphene on top so that to minimize the lattice mismatch to less than 1.5 \%. A sufficient vacuum layer of 20 \AA\ thickness was added to the heterostructure. The atomic coordinates were relaxed until the forces became smaller than 1 meV/\AA. A kinetic energy cutoff of 550 eV has been used for the plane-wave basis set and a $\Gamma$-centered 15$\times$15$\times$1 k-mesh was used to sample the first Brillouin zone. To describe correctly the interaction across the interface, van der Waals forces were used with Grimme type dispersion-corrected density functional theory-D2 \cite{Bucko_2010}. The two-dimensional spin textures were calculated for a 10$\times$10 $\Gamma$-centered 2D k-mesh and the PyProcar package \cite{Herath_2020} was used to plot them. In Fig.~\ref{Fig6}a and ~\ref{Fig6}d, we compare the calculated band structures for a freestanding PtSe$_2$ and that of monolayer-PtSe$_2$/Gr heterostructure. The relaxed interlayer distance between PtSe$_2$ and Gr is found to be 3.29 \AA\ which induces charge transfer across the interface depicted by the charge clouds in the inset of Fig.~\ref{Fig6}d. We estimate a charge transfer of about 0.04 $e$ transferred from the graphene to the PtSe$_2$; a relatively small value that is consistent with the weak vdW interaction type. This is also reflected through the preserved band structure features of pristine graphene and PtSe$_2$ in the heterostructure which is consistent with previous theoretical reports \cite{Sattar_2017}.
To assess the potential Rashba effect at PtSe$_2$/Gr interface, the calculated spin textures for monolayer PtSe$_2$ and PtSe$_2$/Gr heterostructure are compared. Two representative bands are chosen: one in the conduction band (CB) and the other in the valence (VB) that are highlighted in blue and red in Fig.~\ref{Fig6}a, respectively. The corresponding two-dimensional spin textures displayed for the CB (VB) in Fig.~\ref{Fig6}b and ~\ref{Fig6}c, show two contours with opposite spin helicities, however, they are degenerate with no splitting. This is implicit since there is no source of inversion symmetry breaking in a freestanding PtSe$_2$ monolayer. In this respect, the spin textures observed in Refs.\cite{Yao_2016,Yan_2017} can only be explained by the presence of an interface electric field. Interfacing PtSe$_2$ monolayer with graphene breaks the symmetry due to the interfacial dipole and thus induces a Rashba splitting of those bands as can be seen from the spin textures in Fig.~\ref{Fig6}e and ~\ref{Fig6}f. Indeed, the band splitting is found to be larger for the VB with an estimated Rashba parameter $\alpha_R\approx -195$ meV.\AA\ and can be explained by its orbital character which is mainly contributed by the p-orbitals of Se whereas the CB is dominated by Pt d-orbitals. Despite the fact that Pt has larger spin-orbit strength, the Se atoms are positioned at the interface which makes their bands more sensitive to the interfacial dipole effect. Although often neglected in the transition metal systems owing to the reduced hole velocity in the VB, the first step in THz emission (Fig.~\ref{Fig6}e) most probably originates from IREE in this valence band where spin polarized hot holes are injected from CoFeB. However, it is difficult to estimate the fraction of hot holes reaching this band since the position of the Fermi level altered by CoFeB deposition is unknown. ISHE in multilayers PtSe$_2$, corresponding to the second step in THz emission (Fig.~\ref{Fig6}e), can be of extrinsic or intrinsic origin and further investigation would be required to identify its exact mechanism. Nevertheless, as shown in the Supp. Info., we already observe avoided band crossings in the VB of 3 ML PtSe$_2$ that are supposed to contribute to intrinsic ISHE.

\section{Conclusion}
In conclusion, we have realised large area and high quality 2D PtSe$_2$ on graphene by molecular beam epitaxy, on which in-situ grown CoFeB is sputtered to study THz SCC phenomena. By performing a careful and extensive characterization of PtSe$_2$ layers before and after CoFeB deposition, we find that the 2D material retains its structural and chemical properties. For SCC, we performed spintronic THz emission spectroscopy and find efficient THz emission from PtSe$_2$, whereas VSe$_2$ and WSe$_2$ give negligible signals owing to low SOC and 1H crystal symmetry, respectively. From the PtSe$_2$ heterostructures, we clearly demonstrate the magnetic origin of the THz signal and unveil the physical origin of the SCC through a PtSe$_2$ thickness dependence study. The emitted THz electric field as a function of the number of PtSe$_2$ monolayers exhibits a clear two-step behaviour, which reflects the transition from the IREE in semiconducting thin PtSe$_2$ ($\leq$3 ML) to ISHE in thicker semimetallic PtSe$_2$ ($>$3 ML). We also deduce the vertical spin diffusion length in PtSe$_2$ to be 2-3 nm and find that IREE is twice as efficient compared to ISHE for SCC. Our conclusions are supported by first principles calculations showing the existence of Rashba SOC at the PtSe$_2$/Gr interface. The unique band structure flexibility of PtSe$_2$ makes it an ideal candidate as a THz spintronic 2D material and to further explore the underlying mechanisms and engineering of the SCC for THz emission.

\section{Experimental Section}
\textbf{Raman spectroscopy}\\
For Raman measurements, we used a Renishaw INVIA 1 spectrometer with a green laser (532 nm) and a 1800 gr/mm grating. The microscope objective magnification was $\times$100 and the numerical aperture 0.9. The laser power was $<$150 $\mu$W/$\mu$m$^2$. All the spectra correspond to the average of at least three spectra recorded at different locations on the samples.\\
\textbf{X-ray diffraction}\\
The grazing incidence X-ray diffraction (GIXD) was done with a SmartLab Rigaku diffractometer equipped with a copper rotating anode beam tube (K$_\alpha$=1.54 \AA) operating at 45 kV and 200 mA. Parallel in-plane collimators of 0.5° of resolution were used both on the source and detector sides. The out-of-plane XRD measurements were performed using a Panalytical Empyrean diffractometer operated at 35 kV and 50 mA, with a cobalt source, (K$_\alpha$= 1.79 \AA). A PIXcel-3D detector allowed a resolution of 0.02° per pixel, in combination with a divergence slit of 0.125°. Both diffractometers are equipped with multilayer mirror on the incident beam and K$_\beta$ filter on the diffracted beam.\\
\textbf{Transmission electron microscopy}\\
Scanning transmission electron microscopy (STEM) measurements
were performed using a Cs-corrected FEI
Themis at 200 kV. HAADF-STEM (high-angle annular
dark field) images were acquired using a convergence angle
of 20 mrad and collecting electrons scattered at angles
higher than 60 mrad. STEM specimens were prepared by
the focused ion beam (FIB) lift-out technique using a Zeiss
Crossbeam 550. The sample was coated with protective
carbon and platinum layers prior to the FIB cut.\\
\textbf{X-ray photoemission spectroscopy}\\
XPS was performed in situ using a Staib Instruments spectrometer. We used an aluminum anode with $K_{\alpha}$ emission operating at 300 W. The signal was optimized on the Se LMM Auger spectra at a binding energy of 184 eV. The C1s line was used to set the binding energy scale.


\medskip
\section{Acknowledgements} \par 
The authors acknowledge the support from the European Union’s Horizon 2020 research and innovation Programme under grant agreement No 881603 (Graphene Flagship), No 829061 (FET-OPEN NANOPOLY) and No 964735 (FET-OPEN EXTREME-IR). The French National Research Agency (ANR) is acknowledged for its support through the ANR-18-CE24-0007 MAGICVALLEY and ESR/EQUIPEX+ ANR-21-ESRE-0025 2D-MAG projects. The LANEF framework (No. ANR-10-LABX-0051) is acknowledged for its support with mutualized infrastructure.

\medskip

%

\newpage


\newpage

\newpage

\newpage

\newpage

\newpage

\end{document}